\documentclass[a4paper]{jpconf}
\usepackage{graphicx}
\usepackage{amsmath} 
\usepackage{cite}
\usepackage{hyperref}
\usepackage{amsmath,amssymb}
\usepackage{color}
\begin{document}

\title{BTZ black hole assuming running couplings}
\author{\'Angel Rinc\'on, Benjamin Koch, Ignacio Reyes}
\address{Physics Institute, Pontifical Catholic University of Chile, Av. Vicu\~na Mackenna 4860, Santiago, Chile}
\ead{\href{mailto:arrincon@uc.cl}{\nolinkurl{arrincon@uc.cl}}}

\begin{abstract}
In the present work a generalization of the BTZ black hole is studied, for the case of scale dependent couplings. One starts by using the effective action for scale dependence couplings to get a generalization of the Einstein field equations. Self consistent solutions for lapse function, cosmological coupling and Newtons coupling are found. The effect of scale dependent couplings with respect to the classical solution is shown. Moreover, asymptotic behavior as well as thermodynamic properties were investigated. Finally, an alternative way to get the scale dependent Newton coupling, from the so-called ``Null Energy Condition" is presented.
\end{abstract}

\section{Introduction}
Great effort has been made to try to unify general relativity with quantum mechanics. 
Two well known candidates for this unification are String theory \cite{Polchinski:1998rq,Polchinski:1998rr} and Loop quantum gravity \cite{book:217893}. Another promising framework for quantum gravity arises from the so called Asymptotic safety scenario
\cite{book:112336,Reuter:2012id,Weinberg:1979,Wetterich:1992yh,Dou:1997fg,Reuter:2001ag}, in which the couplings do not need to be small or tend to zero in the high energy limit.
For this scenario to work it necessary to have a finite number of couplings that have a stable UV fixed point.
Evidence for the existence of such a non-trivial UV fixed point has been found
by the use of non-perturbative RG equations \cite{lrr-2006-5}. 
\noindent Independent of the particular RG approach, the outcome of such calculations will give an effective gravitational theory with scale dependence couplings. 
As simpler toy model for gravity with scale dependent couplings we study three dimensional gravity, which allows for many non-trivial
effects despite of the fact that there are no classical propagating degrees of freedom.
For example it is known that this theory has a close connection with Chern-Simons theory \cite{Witten:2007kt,Witten:1988hc}.
Another particularly interesting feature of this theory is that there are non-trivial black hole solutions with a negative cosmological constant, found 
by Ba\~nados, Teitelboim and Zanelli (BTZ) \cite{Banados:1992wn,Banados:1992gq}.
We investigate this black hole in light of the  possibility of scale dependent couplings, 
such as they arise in the asymptotic safety approach. 
\noindent Using the Einstein-Hilbert action, where the couplings are allowed to vary with respect to an arbitrary scale, we derive generalized field equations for the metric and the scale field. Solutions for those equations are obtained by using the schwarzschild ansatz.

\noindent The paper is organized as follows: In Section \ref{Main_Idea} the fundamental ideas, as well as the required framework are discussed. In section \ref{Results} the main results are collected. The consequences of our solution are discussed and collected in sections \ref{Discussion} and \ref{Conclusion}.

\section{Main idea}\label{Main_Idea}
\noindent Pioneer works in which quantum gravity is investigated via the effective average action are found at Ref. \cite{Reuter:1996cp,Reuter:2003ca}. 
\noindent In this work, the asymptotic safety ideas previously commented are used. Starting from the Einstein Hilbert action with cosmological constant one take advantage of the Einstein field equation to obtain the equation of motion (hereafter E.O.M.). Thus the effective action, independent of the kind of matter, is
\begin{align}
\Gamma_k[g_{\mu \nu},k] &= \int d^3 x \sqrt{-g}\Bigg[ \frac{R - 2 \Lambda_k }{G_k}\Bigg],
\end{align} 
one obtains
\begin{align}\label{EOM}
G_{\mu \nu} + \Lambda_k g_{\mu \nu} &=  8\pi G_k T_{\mu \nu}, 
\\
R \frac{\partial}{\partial k} \left(\frac{1}{G_k}\right)  &= 2 \frac{\partial}{\partial k}\left( \frac{\Lambda_k}{G_k}\right), 
\end{align}
where the effective energy-momentum tensor is given by       
\begin{align}
 8\pi G_k T_{\mu \nu} &=   8\pi G_k T^m_{\mu \nu}  - \Delta t_{\mu \nu}.
\end{align}
Note that $T^{m}_{\mu \nu}$ is the energy-momentum tensor associated with matter and $\Delta t_{\mu \nu}$ reads
\begin{align}
\Delta t_{\mu \nu} &= G_{
k}\Bigl(g_{\mu\nu} \square -\nabla_\mu\nabla_\nu \Bigl)G_{k}^{-1}.
\end{align}
Please, note that $k$ is an arbitrary renormalization 
scale. If one is interested in purely spherically symmetric settings, one
knows that $\mathcal{Q} \equiv \mathcal{Q}\big(k(r)\big) \rightarrow \mathcal{Q}(r)$, where $\mathcal{Q}$  symbolizes 
unknown functions of the problem e.g. $\{f(r), \Lambda(r), G(r)\}$. 
Thus, one only needs to solve the E.O.M. for the radial coordinate $r$ \cite{Koch:2010nn,Koch:2015nva,Koch:2014joa}, if one chooses to eliminate one of those functions by a suitable ansatz or a physically motivated condition.

\section{Results}\label{Results}
One starts by considering a metric in (2+1) dimensional case
\begin{eqnarray}
ds^2= -f(r) \, dt^2+ g(r) \, dr^2 + r^2 d\phi^2 ,
\end{eqnarray}
where $f(r)$ is the lapse function. For vanishing angular momentum ($J_0 = 0$), the so-called ``Null Energy Condition" \cite{Rubakov:2014jja} gives a relation between $f(r)$ and $g(r)$ such that $g(r)=1/f(r)$. This condition is inspired by the Jacobson argument in which one assumes a specific null vector $\ell^{\mu}$ with $R_{\mu \nu} \ell^{\mu} \ell^{\nu} = 0$ \cite{Jacobson:2007tj}. Moreover, according with the E.O.M. (\ref{EOM}) one sees that 
\begin{align}
\Delta t_{\mu \nu}\ell^{\mu} \ell^{\nu} = 0, \label{Deltat}
\end{align}
which allows to find gravitational scale dependent coupling without actually solve (\ref{EOM})
\begin{equation}
G(r) = \frac{G_0^2}{G_0 + (1 + G_0 M_0)\epsilon r}. \label{Gsol}
\end{equation}
Here the constants $\{G_0, M_0, \epsilon\}$ were chosen such that (\ref{Gsol}) coincides with the classical BTZ solution for $J_0 = 0$. Defining $\delta(r)$ as a contrast to Newtons constant
\begin{equation}\label{delta}
\delta(r) =  \frac{G_0}{G(r)} -1,
\end{equation}
one can express the generalized solutions for $f(r)$ and $\Lambda(r)$ using (\ref{delta}) and solving the Eq. (\ref{EOM})
\begin{align} \label{fAngel}
f(r) &=  f_{0}(r) + 2 M_0 G_0 \delta(r) 
\left[
1 + \delta(r) \ln \left(\frac{\delta(r)}{1+\delta(r)}\right)
\right], \\
\Lambda(r) &=  - \frac{1}{\ell_0 ^2}
\left(\frac{1}{1+\delta(r)}\right)^2
\Bigg[ 1 +4 \delta(r) +
\left(5 M_0 G_0 \frac{\ell_0^2}{r^2}  +3\right) \delta(r) ^2 + 
6 M_0 G_0 \frac{\ell_0^2}{r^2} \delta(r)^3 
\nonumber
\\
& \ \ \
+ 
2 M_0 G_0 \frac{\ell_0^2}{r^2}
\Bigl(1 + \delta(r)\Bigl)
\Bigl(3 \delta(r) +1 \Bigl) \delta(r)^2 \ln \left(\frac{\delta(r)}{1+\delta(r)}\right)\Bigg].
\end{align}
The set of solutions $ \{G(r), f(r), \Lambda(r)\}$ could be either expressed in terms of $\epsilon$ parameter (which measures the deviation from the classical solution) or, alternatively, in terms of $\delta(r)$ which measures the same effect but in a more intuitive way. Thus, when $\delta(r) = 0$ the classical case is recovered, and when $\delta(r) \neq 0$ improved solutions (which show the effect of scale dependent couplings) are obtained. In Fig. \ref{fr} the lapse function $f(r)$ is shown for different values of $\epsilon$.

\subsection{Horizons}
\begin{figure}[h]
\begin{minipage}{17pc}
\includegraphics[width=17pc]{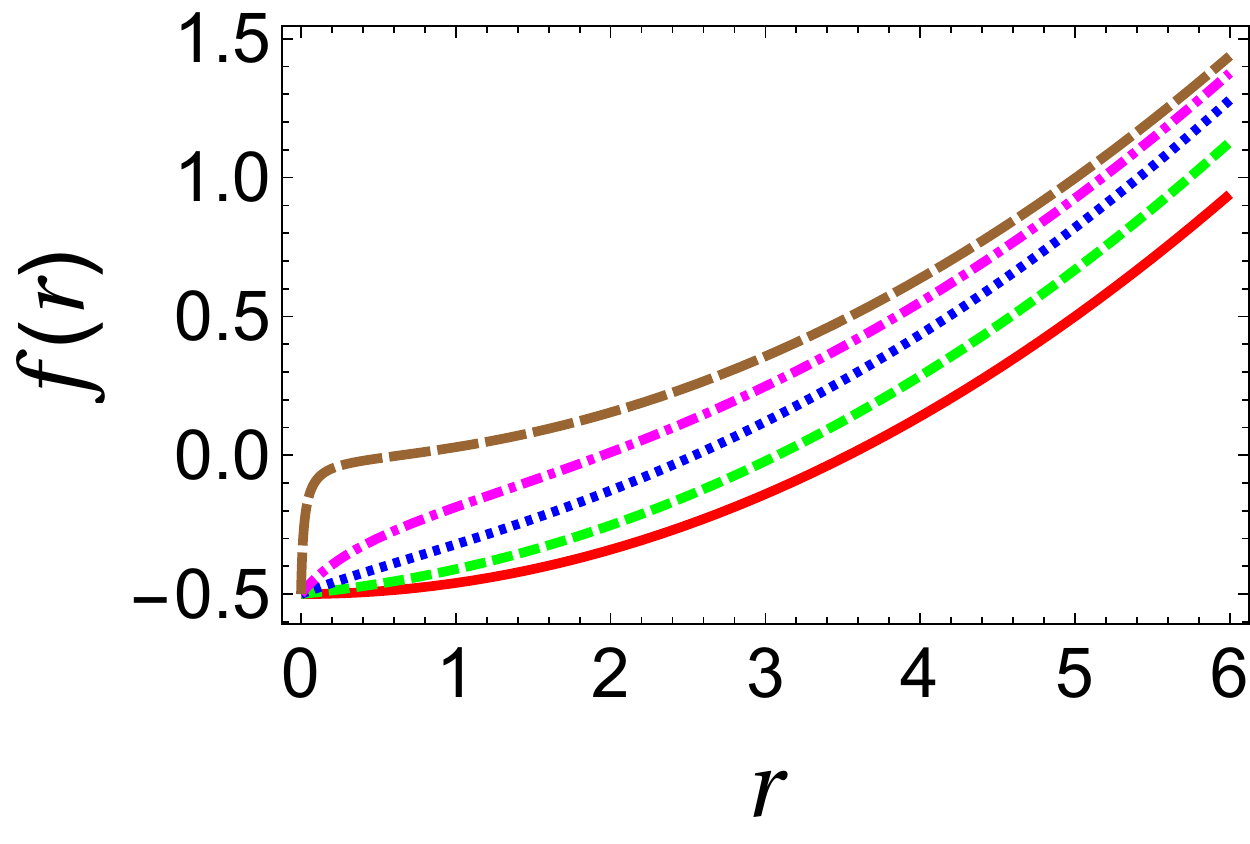}
\caption{\label{fr} Radial dependence of the lapse function $f(r)$ for $\ell_0=5$, $G_0=1$ and $M_0=0.5$. The different curves correspond to the classical case $\epsilon=0$ 
solid red line, 
$\epsilon=0.04$ 
short dashed green line,
$\epsilon=0.15$ 
dotted blue line, 
$\epsilon=0.5$ 
dot-dashed magenta line, and 
$\epsilon=20$ 
long dashed brown line.}
\end{minipage}\hspace{2pc}%
\begin{minipage}{17pc}
\includegraphics[width=17pc]{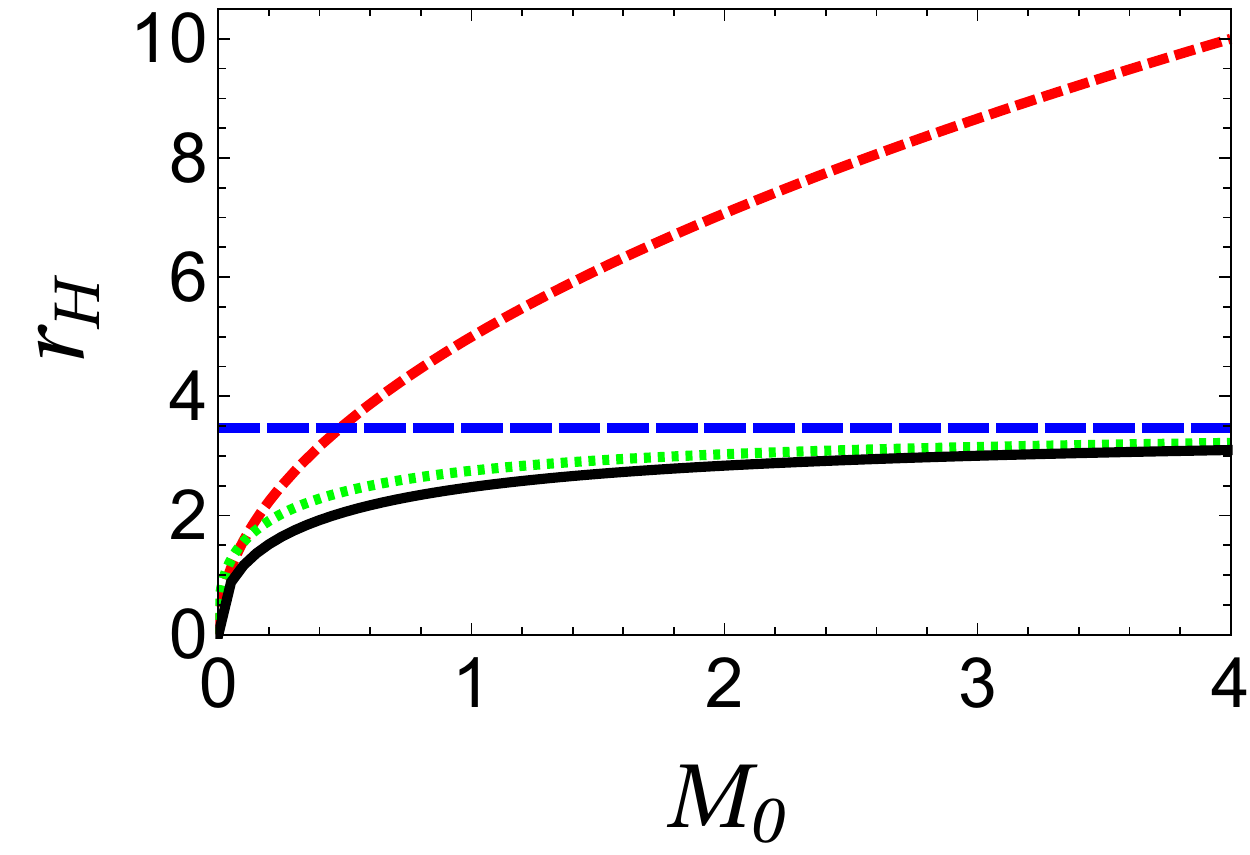}
\caption{\label{rh}
Horizon $r_H$ as a function of $M_0$ for $\ell_0=5$, $G_0=1$ and $\epsilon = 0.4$.
The curves correspond to the classical case
(short dashed red line)
for small $G/G_0$ 
(dotted green line),
for small $G/G_0$ and large $G_0 M_0$ 
(long dashed blue line),
and the numerical solution 
(solid black line).
}
\end{minipage} 
\end{figure}

In order to have a well-defined black hole surface, a horizon must exist. Horizons are crucial for the understanding of the thermodynamics properties of the black hole. The aparent horizon is found as solution of
$f(r_H)=0$. An analytic expression for the case of $r_H$ can not be obtained. However in certain limits such as $\epsilon \ll 1$ or when $G(r)/G_0$ is small and, simultaneously $M_0 G_0$ is large, analytic expressions are found. Check table (\ref{ex}) to see details. In Fig. (\ref{rh}) one can see the scale dependent effect on the horizon structure of the generalized BTZ solution.

\subsection{Thermodynamics}
\begin{center}
\begin{figure}[h]
\begin{minipage}{17pc}
\includegraphics[width=17pc]{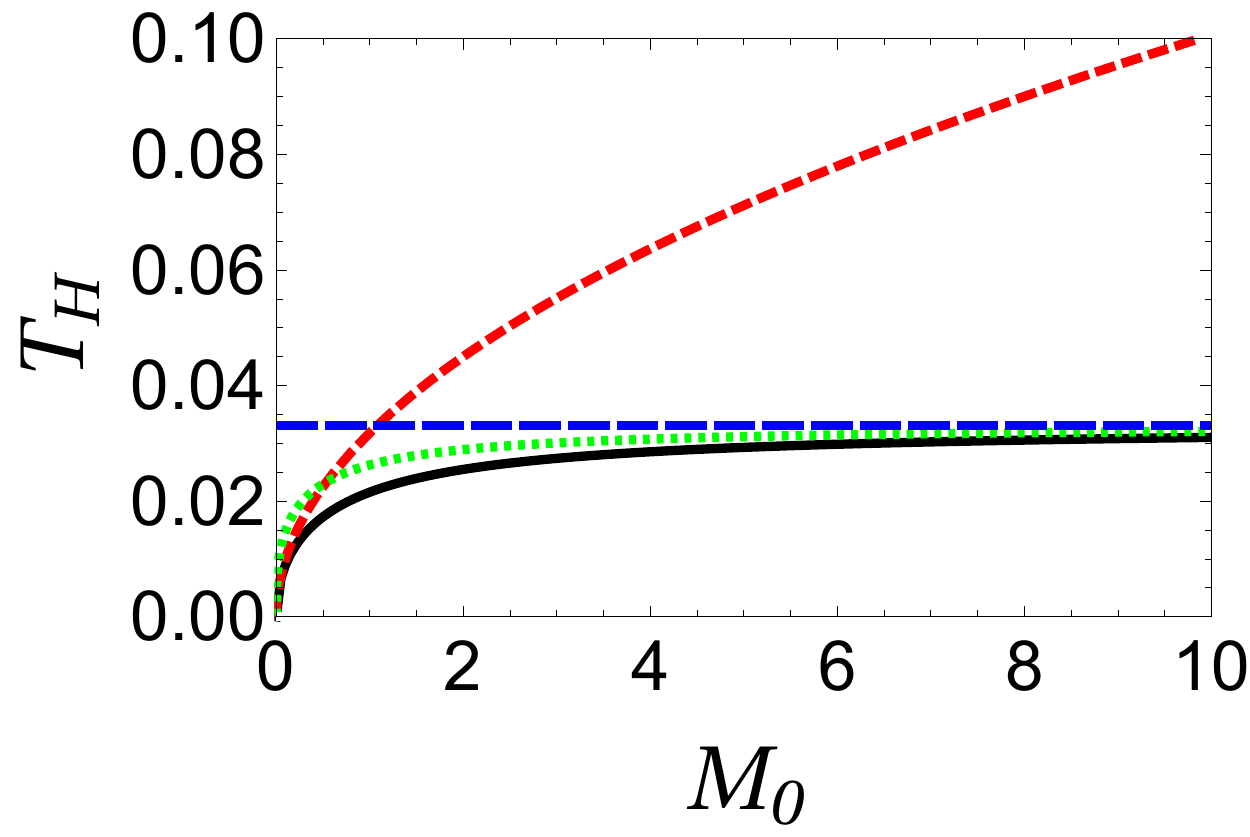}
\caption{\label{Th}
Temperature $T_H $ as a function of $M_0$ for $\ell_0=5$, $G_0=1$, and $\epsilon = 0.4$.
The curves correspond to the classical case 
(short dashed red line),
for small $G/G_0$ 
(dotted green line),
for small $G/G_0$ and large $G_0 M_0$ 
(long dashed blue line),
and the numerical solution 
(solid black line).
}
\end{minipage}\hspace{2pc}%
\begin{minipage}{17pc}
\includegraphics[width=17pc]{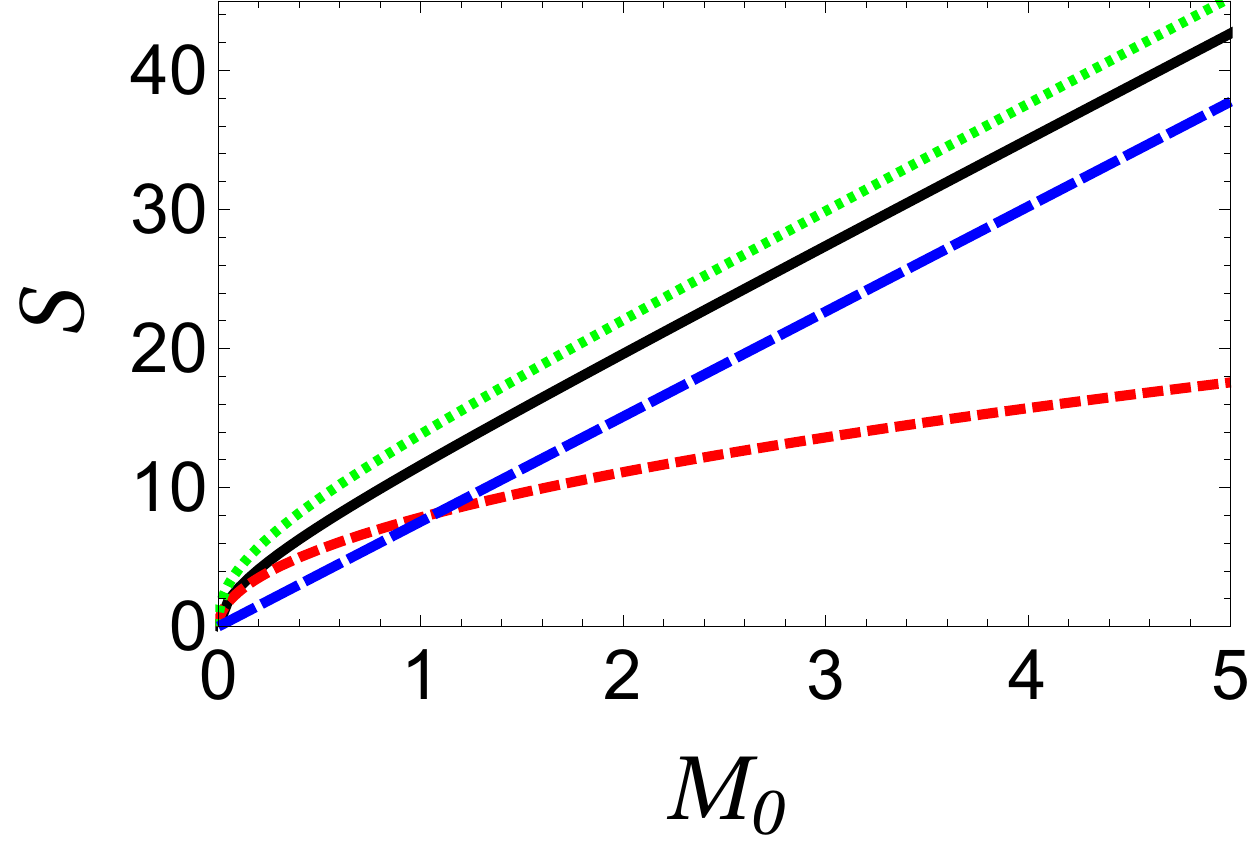}
\caption{\label{S}
Entropy $S$ as a function of $M_0$ for $\ell_0=5$, $G_0=1$, and $\epsilon = 0.4$. 
The curves correspond to the classical case 
(short dashed red line),
for small $G/G_0$ 
(dotted green line),
for small $G/G_0$ and large $G_0 M_0$ 
(long dashed blue line),
and the numerical solution 
(solid black line).
}
\end{minipage}
\end{figure}
\end{center}

The Bekenstein-Hawking (BH) entropy as well as Hawking temperature are revisited. The usual relation for BH entropy is $S_0=A/4 G$. In the scale dependent case this relation reads \cite{Jacobson:1993vj,Iyer:1995kg,Visser:1993nu,Creighton:1995au,Kang:1996rj}
\begin{align}
S &= S_0 \cdot \left[ 1 + 
\frac{(1 + M_0 G_0)\epsilon r}{G_0}\right]\Bigg{|}_{r_H}.
\end{align} 
The Hawking temperature is obtained by the usual relation $T_H = f'(r_H)/4\pi$. This condition remains even for running couplings. However, since the horizons experience a shift due to $\epsilon \neq 0$, also the Hawking temperature does,
\begin{align}
T_H &=\left.\frac{1}{2 \pi r}\frac{G_0^2 M_0}{G_0+ (1+G_0 M_0)\epsilon r }\right|_{r_H}.
\end{align} 
The running coupling effect on $S$ and $T_H$ is shown in Fig. (\ref{Th}) and Fig. (\ref{S}). 
As mentioned before, those numerical results can be approximate analytically in certain regimens which are summarized in the Table \ref{ex}. Finally, further details about this solution as well as alternative discussion could be found in Ref. \cite{Koch:2016uso}.

\begin{table}[h]
\caption{\label{ex} Black hole relevant quantities at different approximations.}
\begin{center}
\begin{tabular}{llll}
\br
Approximation & Entropy $S$ & Temperature $T_H$ & Horizon $r_H$\\
\mr
$\epsilon \ll 1$ &
$\frac{\pi}{2}\sqrt{\frac{M_0}{G_0}}\ell_0$  &
$\frac{\sqrt{M_0G_0}}{2 \pi \ell_0}$ &
$ \sqrt{G_0 M_0} \ell_0$
\\
\\
$G(r)/G_0 \ll 1$  & 
$\pi  \bigg[  \frac{\ell_0 ^4 M_0 ^2 (1 + M_0 G_0)\epsilon}{18 G_0 ^2} \bigg]^{\frac{1}{3}}$ &
$ \frac{1}{4\pi} \bigg[ 18 \frac{M_0G_0^2}{\ell_0^4(1+G_0M_0)\epsilon}  \bigg]^{\frac{1}{3}}$ &
$\bigg[ \frac{2}{3}\frac{M_0 G_0^2\ell_0^2}{(1+M_0 G_0)\epsilon}\bigg]^{\frac{1}{3}}$
\\
\\
$
\begin{array}{lcl}
\hspace{-0.18cm}G(r)/G_0 \ll 1 
\\
\hspace{-0.18cm}M_0G_0 \gg 1
\end{array}
$&
$\pi M_0 \bigg[\frac{\ell_0^4 \epsilon}{18 G_0}\bigg]^{\frac{1}{3}}$ &
$\frac{1}{4\pi} \bigg[ 18 \frac{G_0}{\ell_0^4\epsilon} \bigg]^{\frac{1}{3}}$&
$\bigg[ \frac{2}{3}\frac{G_0 \ell_0^2}{\epsilon}\bigg]^{\frac{1}{3}}$
\\
\br
\end{tabular}
\end{center}
\end{table}

\section{Discussion}\label{Discussion}
Due the logarithmic contribution in Eq. (\ref{fAngel}) it is impossible to find an exact expression for the horizon. This logarithmic term is induced by the scale dependent effect associated with the additional object $\Delta t_{\mu \nu}$. However, it is possible to show that the scale dependent framework does not introduce a new horizon and, therefore, only one horizon appears. The BH entropy grows monotomously with $M_0$. 
In addition, the temperature decreases when $M_0$ goes to zero and tends to a finite value when $M_0$ goes to infinity.

\section{Conclusion}\label{Conclusion}

To summarize, in this article the BTZ black hole is investigated in the light of scale dependence. Analytical expressions for the lapse function, the cosmological coupling and, the Newton coupling are determined by solving the Einstein field equations. In figures \ref{fr}, \ref{rh}, \ref{Th} and \ref{S}, the scale dependent effect on the classical BTZ black hole is shown. Due the scale dependence, there appears an effective energy momentum tensor $\Delta t_{\mu \nu}$ associated with the running Newton coupling.
According to the ``null energy condition'' (which reads $R_{\mu \nu} \ell^{\mu} \ell^{\nu} = 0$) the usual ansatz  $g_{tt}=-1/g_{rr}\equiv f(r)$ is derived i.e. the so called ``Schwarzschild relation'' is preserved. 
Special attention is dedicated to the interpretation of the integration constants
which is given in terms of the classical parameters $\{G_0$, $\ell_0$, $M_0\}$
and one additional constant $\epsilon$, that parametrizes the strength of
scale dependence.
Horizons, as well as black hole thermodynamics, are investigated. It is found that the large $r$ asymptotic is $AdS_3$ and that the $r\rightarrow 0$ asymptotic has a singular behaviour, not present in the usual BTZ solution. Furthermore, it is found that for fixed values of $\{\epsilon,\, G_0, \ell_0\}$ the horizon radius saturates  for $M_0 \rightarrow \infty$ to a finite value.
Finally, an important modification of the classical result is found: when $G(r)$ deviates strongly from $G_0$ a transition from the standard relation $S_0= A/4G$ to $S= S_0 \times (1 + \alpha r)$ is found. 
This apparent deviation from the holographic principle
is an unusual feature of this new black hole solution using scale dependent couplings.

\section*{Acknowledgments}
The author A.R. was supported by the CONICYT-PCHA/Doctorado Nacional/2015-21151658.
The work of B.K. was supported by the Fondecyt 1161150 and the work of I.R. was partially funded by CONICYT PCHA/Doctorado Nacional \# 2015149744.

\section*{References}
\bibliographystyle{iopart-num}
\bibliography{refsBTZ,myproyect}

\end{document}